\input{aipcheck}

\documentclass[
  ]
  {aipproc}

\layoutstyle{6x9}

\def\met{\mbox{\ensuremath{\, \slash\kern-.6emE_{T}}}}
\def\TeV{\ifmmode {\mathrm{\ Te\kern -0.1em V}}\else
                   \textrm{Te\kern -0.1em V}\fi \xspace}%
\def\GeV{\ifmmode {\mathrm{\ Ge\kern -0.1em V}}\else
                   \textrm{Ge\kern -0.1em V}\fi \xspace}%

\def\stop{\ensuremath{\tilde{t}} \xspace}

\def\chinopm{\ensuremath{\mathchoice%
      {\displaystyle\raise.4ex\hbox{$\displaystyle\tilde\chi^\pm$}}%
         {\textstyle\raise.4ex\hbox{$\textstyle\tilde\chi^\pm$}}%
       {\scriptstyle\raise.3ex\hbox{$\scriptstyle\tilde\chi^\pm$}}%
 {\scriptscriptstyle\raise.3ex\hbox{$\scriptscriptstyle\tilde\chi^\pm$}}} \xspace}
\def\nino{\ensuremath{\mathchoice%
      {\displaystyle\raise.4ex\hbox{$\displaystyle\tilde\chi^0$}}%
         {\textstyle\raise.4ex\hbox{$\textstyle\tilde\chi^0$}}%
       {\scriptstyle\raise.3ex\hbox{$\scriptstyle\tilde\chi^0$}}%
 {\scriptscriptstyle\raise.3ex\hbox{$\scriptscriptstyle\tilde\chi^0$}}} \xspace}
\def\ninoone{\ensuremath{\mathchoice%
      {\displaystyle\raise.4ex\hbox{$\displaystyle\tilde\chi^0_1$}}%
         {\textstyle\raise.4ex\hbox{$\textstyle\tilde\chi^0_1$}}%
       {\scriptstyle\raise.3ex\hbox{$\scriptstyle\tilde\chi^0_1$}}%
 {\scriptscriptstyle\raise.3ex\hbox{$\scriptscriptstyle\tilde\chi^0_1$}}} \xspace}
\def\ninotwo{\ensuremath{\mathchoice%
      {\displaystyle\raise.4ex\hbox{$\displaystyle\tilde\chi^0_2$}}%
         {\textstyle\raise.4ex\hbox{$\textstyle\tilde\chi^0_2$}}%
       {\scriptstyle\raise.3ex\hbox{$\scriptstyle\tilde\chi^0_2$}}%
 {\scriptscriptstyle\raise.3ex\hbox{$\scriptscriptstyle\tilde\chi^0_2$}}} \xspace}

\begin{document}

\title{Prospects for SUSY searches in CMS and ATLAS}



\author{Paul de Jong, on behalf of the CMS and ATLAS collaborations}{
  address={Nikhef, P.O. Box 41882, NL-1009 DB Amsterdam, the Netherlands}
}



\begin{abstract}
 We discuss how the CMS and ATLAS experiments are preparing
 for the analysis of first LHC data with emphasis
 on the search for supersymmetry. We will show the importance of the
 understanding of detector, trigger, reconstruction and backgrounds,
 and we will present realistic estimates of the reach of
 CMS and ATLAS.
\end{abstract}

\maketitle


\section{Introduction}

After a run at $\sqrt{s} =$ 10 TeV in 2008, the Large Hadron Collider
(LHC) at CERN is likely to deliver a few fb$^{-1}$ of integrated
luminosity at 14 TeV in 2009, and keep increasing its luminosity in further
years. The CMS and ATLAS experiments are general multi-purpose
detectors designed to analyze the results of the collisions. Due to
the high centre-of-mass energy and the high luminosity of the LHC, the prospects
for searches for new physics beyond the Standard Model (SM), including
supersymmetry (SUSY) are excellent.

At the time of writing of these proceedings, first beams have been
injected in the LHC.  Both experiments are basically ready for beam;
in the last year(s) both CMS~\cite{cmstdr} and ATLAS~\cite{atlascsc} 
have actively prepared for SUSY searches.  For lack of space, only
a selection of results is shown. Since at the time of writing there
is no LHC data yet, all ``results'' have been obtained with realistic Monte
Carlo simulations. Although this note focuses on searches for supersymmetry,
CMS and ATLAS perform many more searches for general new physics beyond
the SM.

\section{Confidence building}

{\em Searching for SUSY equals confidence building}. Only with confidence
in the operation and performance of the detector, in the trigger, in the
reconstruction and object identification methods, and in our knowledge
of the backgrounds, can we claim to see excesses over the SM and
evidence for new physics. This confidence building will be the
primary activity of the experiments with first data, and it will
need to continue thereafter. 
Although with a very small data sample we are in principle sensitive to SUSY
beyond the Tevatron, confidence building needs time, hard work and luminosity.

The primary objective of the experiments now is to establish
reliable, long-term, controlled and safe running of the detectors.
With systems as complex as those of CMS and ATLAS, this is not
a triviality. During this process, we will need to get to know the
detectors like the back of our hand: their problematic regions,
dead or noisy cells, etc.

Triggering on SUSY should not be difficult: there are likely to
be energetic jets, significant missing transverse energy, electrons,
muons, taus, photons, and/or b-quark jets. In combination, trigger
efficiencies of better than 95\% should be attainable. However, the
trigger performance should be demonstrated from the data, and
the trigger should be designed with this in mind. Also background
control samples should be triggered on.

Confidence in the reconstruction and object (electrons, muons,
jets etc.) identification will be gained from detailed studies of first
data. Calibration- and alignment constants must
be derived and applied, and energy- and momentum scales set.
In 1 pb$^{-1}$ (3 days at $10^{31}$ cm$^{-2}$ s$^{-1}$), 16000 
$J/\Psi \to \mu^+ \mu^-$ and 3000 $\Upsilon \to \mu^+ \mu^-$ events
are expected; in 10 pb$^{-1}$ some 6000 Z $\to \mu^+ \mu^-$ and a simular number
to $e^+ e^-$ should be collected. 
From these relatively simple topologies we move to more
complex ones by allowing additional jets; when we approach 100 pb$^{-1}$ the
top quark becomes an excellent calibration tool. Clean $t \bar{t}$ samples
can be selected and used to verify lepton identification, jet- and
missing transverse energy scales, and b-quark tagging. 

\section{Backgrounds}

Confidence in the understanding of the SM backgrounds can and will be 
gained by a dedicated campaign that involves data-driven background 
determinations, aided by Monte Carlo wherever needed~\cite{legger}. 
The validation of Monte Carlo's with data is also an early goal. A number of 
background estimation methods are under development in CMS and ATLAS;
only their consistency in combination can lead to the
desired confidence in understanding. Even then, statistical and
systematic uncertainties on the backgrounds remain, which must be taken
into account in the SUSY searches.

\begin{figure}[tb]
  \includegraphics[width=0.42\textwidth]{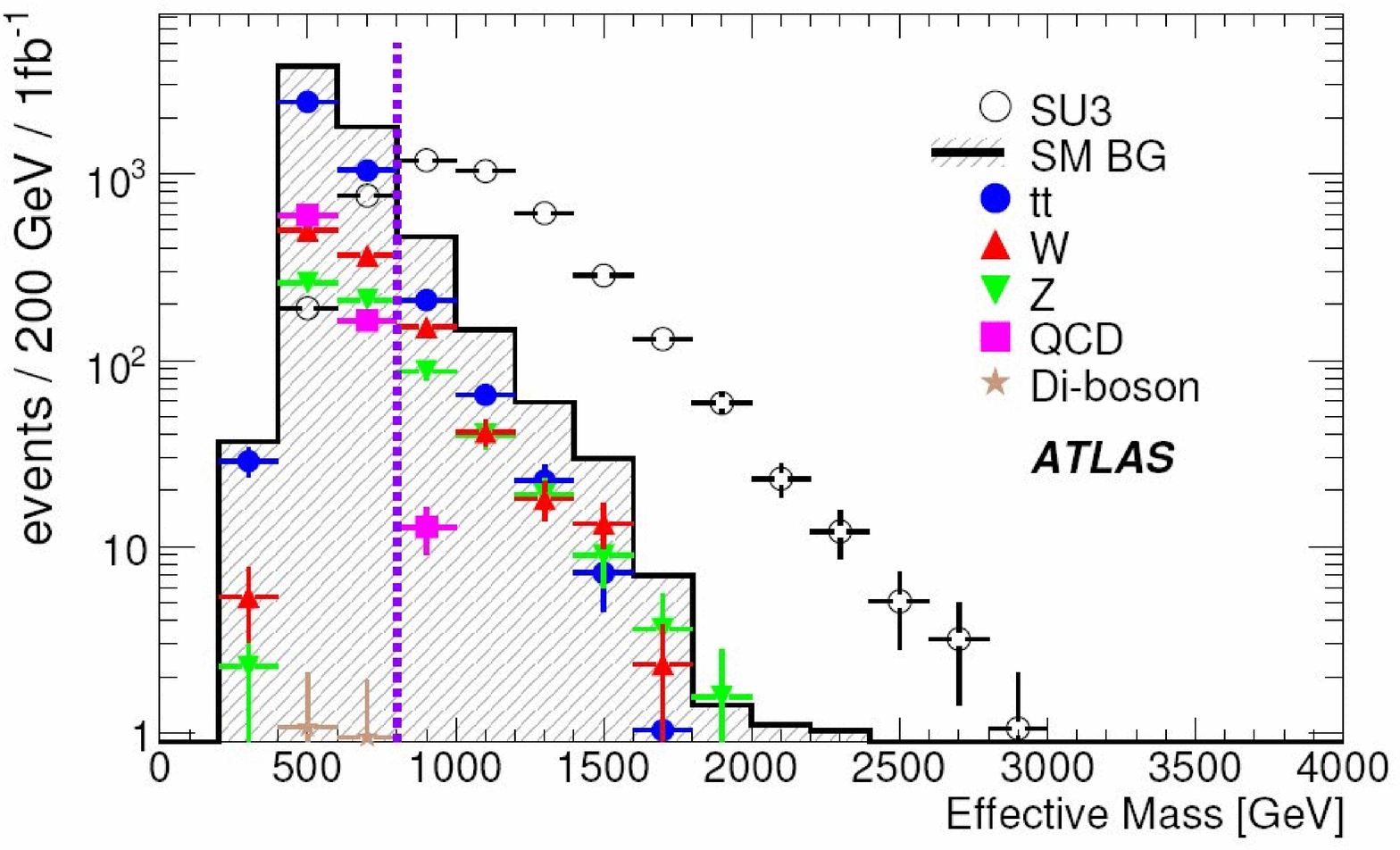}
  \includegraphics[width=0.42\textwidth]{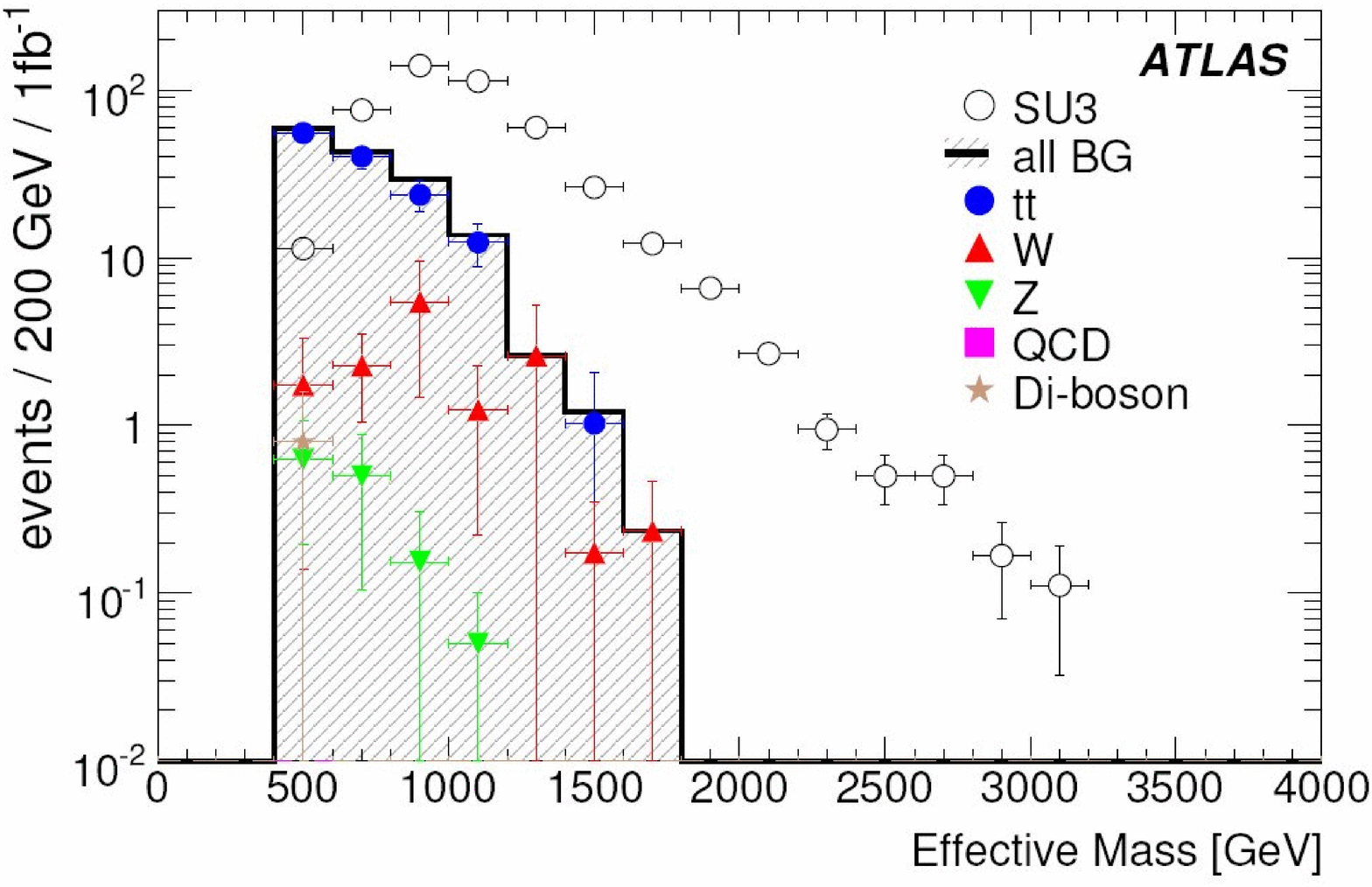} 
  \caption{Backgrounds in the no-lepton (left) and in the one-lepton
   (right) search modes, for ATLAS in 1 fb$^{-1}$, after preselection
   cuts requiring at least four energetic jets and significant missing
   transverse energy. Also the SUSY signal of the SU3 benchmark
   point is shown.
  \label{fig:backgrounds}
}
\end{figure}

After demanding a few energetic jets and some missing transverse energy
$E_T^{\mathrm{miss}}$, as one would do in a preselection for a typical 
R-parity conserving SUSY analysis, the SM background
consists largely of QCD multi-jet production, W and Z production with
extra jets, top-pair production, and to a lesser extent di-boson and
single top production, as shown in Figure~\ref{fig:backgrounds}. 

\begin{figure}[tb]
  \includegraphics[width=0.38\textwidth]{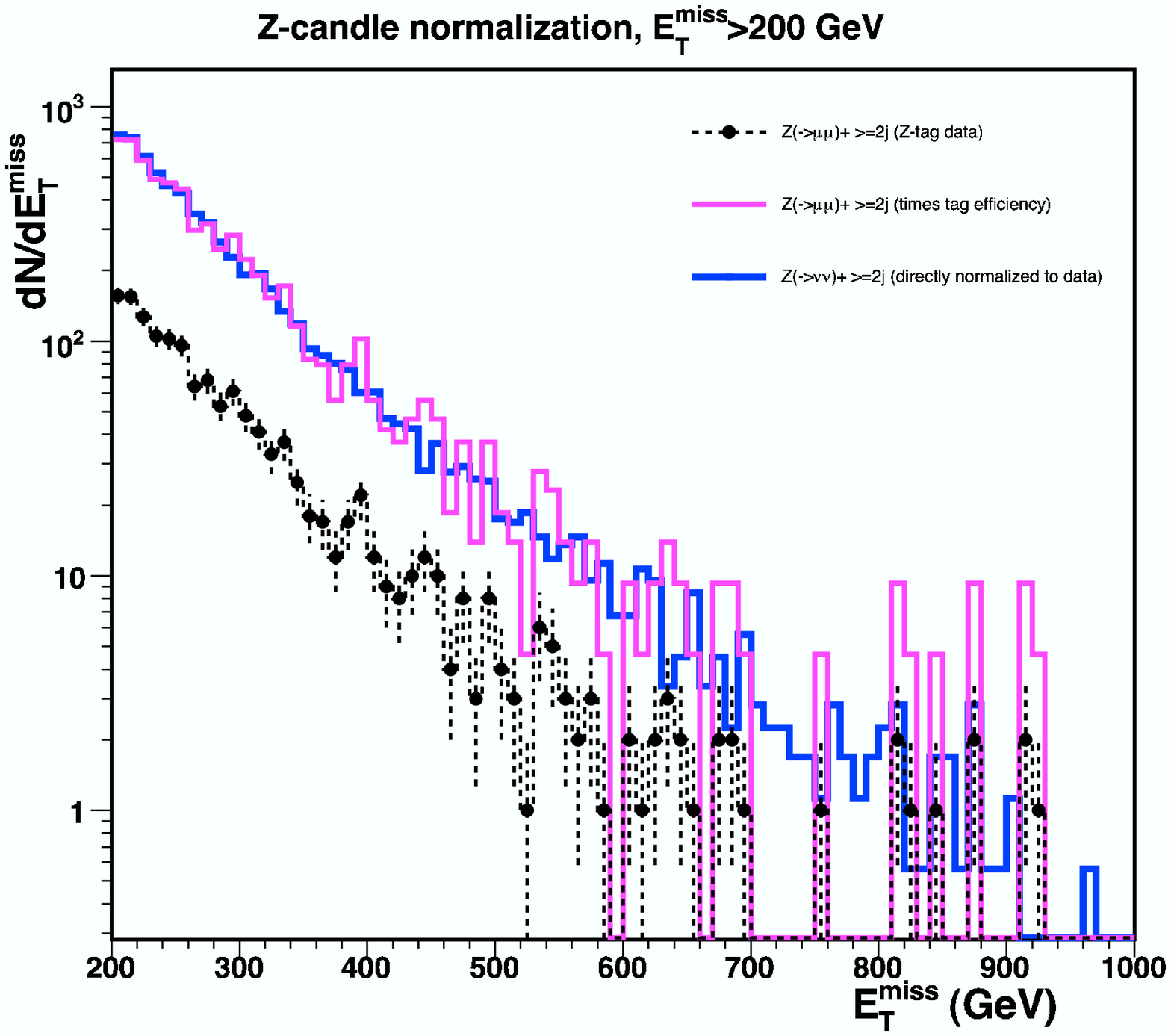}
  \includegraphics[width=0.45\textwidth]{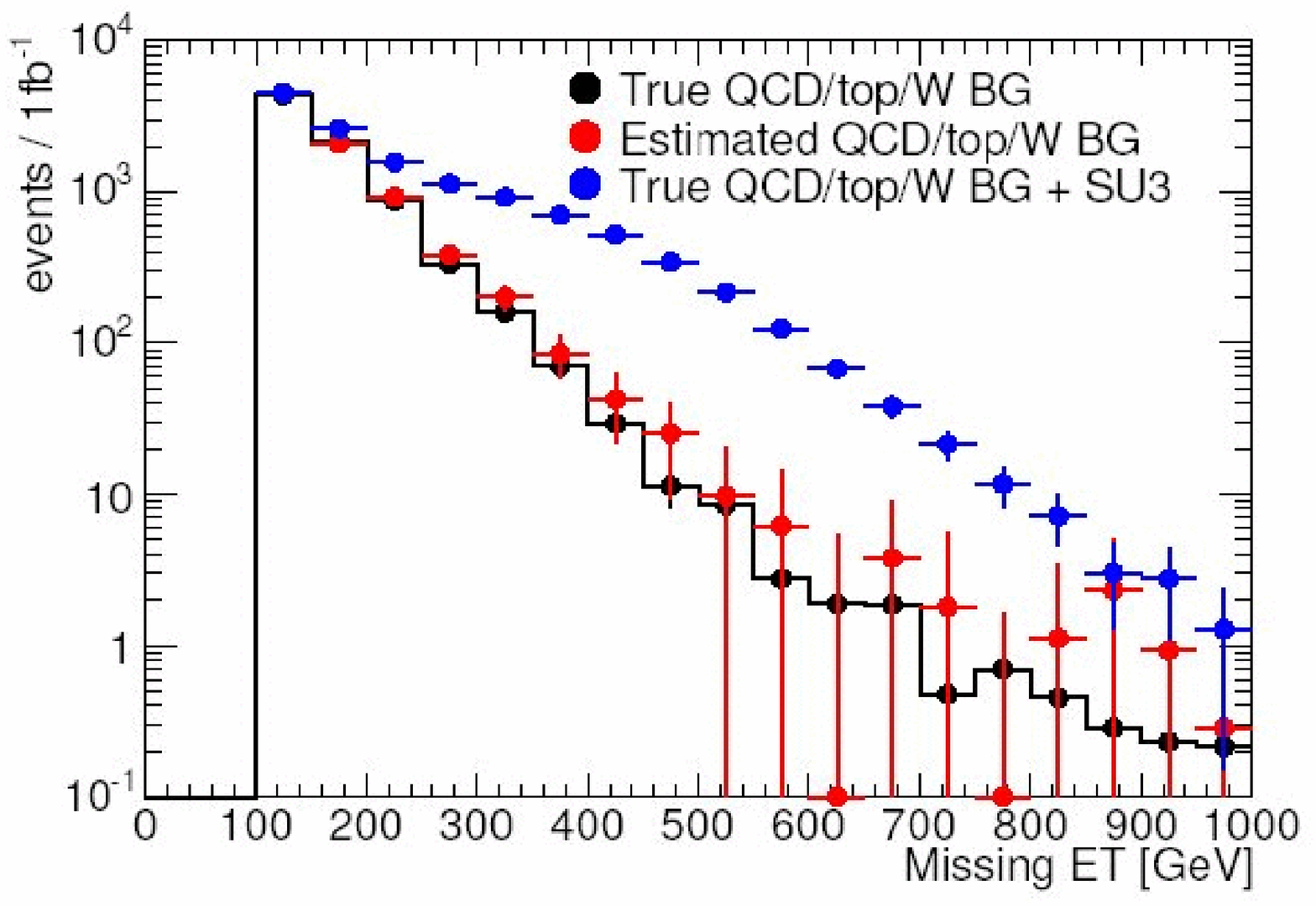}
  \caption{Data-driven methods for background estimation. Left:
  CMS estimate of Z($\rightarrow \nu \nu$)+jets background from
  a Z($\rightarrow \mu^+ \mu^-$)+jets control sample. Right: ATLAS
  estimate of QCD + top + W background in the no-lepton mode,
  compared to the ``true'' background from Monte Carlo, and the
  SU3 SUSY signal.
  \label{fig:datadriven}
}
\end{figure}

\subsubsection{No-lepton final state}


In this search mode, with a veto on isolated leptons, QCD multi-jet production
in principle dominates the background. $E_T^{\mathrm{miss}}$
can be generated by neutrinos from heavy quark decay, or is ``fake'', i.e.
generated by detector effects or backgrounds
unrelated to the collision such as cosmic rays or beam halo. Clean-up
cuts such as calorimeter timing cuts, a good primary vertex, and jet shower
shape cuts are needed.
Mis-measured jets can lead to fake $E_T^{\mathrm{miss}}$; in this
case the $E_T^{\mathrm{miss}}$ direction will point to one of the jets,
and such topologies can be cut away by demanding isolation of the
$E_T^{\mathrm{miss}}$ vector. The remaining $E_T^{\mathrm{miss}}$ in
the QCD multi-jet background should be determined from data. This can be
done using prescaled jet triggers, or with clean samples of top-quark
pairs and Z+jets events. ATLAS has also studied a method that involves 
measuring an $E_T^{\mathrm{miss}}$ response function from data 
(the Gaussian part from photon+jets events, the tails from a sample of 
three-jet events with the missing momentum vector pointing towards or away 
from one of the jets), and applying this function
to a large sample of balanced, low $E_T^{\mathrm{miss}}$, events. 

Further backgrounds in the no-lepton mode include Z+jets events with the Z decaying to
neutrinos, top-quark pair events and W+jets events.
The Z$\rightarrow \nu \nu$ background can be effectively estimated from
Z$\rightarrow \mu^+ \mu^-$ control samples, as shown in 
Figure~\ref{fig:datadriven} (left).
The top and W backgrounds are most dangerous when the W decays into a
lepton and a neutrino, but the lepton is not identified, for example because
it is a tau lepton, or falls outside the acceptance. Various methods
to estimate this from data are being developed, Figure~\ref{fig:datadriven}
(right) shows an example.

\subsubsection{Final state with one lepton}

In the one-lepton search mode, the presence of a high $p_T$ isolated
electron or muon is required. This facilitates triggering on the event
and suppresses the QCD background, but also costs signal efficiency.
Nevertheless, the one-lepton mode is a robust way to look for SUSY and
will play an important role.
Figure~\ref{fig:backgrounds} (right) shows the expected backgrounds for
the one-lepton mode after preselection cuts. It is dominated by top-quark
pairs and W+jets events; a sizable fraction of the top background comes
from dileptonic top events with one lepton not identified. Details of
data-driven estimation methods are described elsewhere in these
proceedings~\cite{legger}.



\subsubsection{Uncertainties on the background}

Table~\ref{tab:bg} lists the uncertainties on the background
as currently expected by ATLAS with data-driven techniques, in 1 fb$^{-1}$.
The statistical uncertainties apply to the ATLAS cuts 
before the final $M_{\mathrm{eff}}$ cut.

\begin{table}
\begin{tabular}{|l|cc|}
\hline
Source & Stat. unc. (\%) & Syst. unc. (\%) \\
\hline
QCD multi-jets  & $1$  &  $50$ \\
top $\to \tau$  & $6$  &  $15$ \\
Z $\to \nu \bar{\nu}$ & $8-13$ & $10-15$ \\ \hline
$t \bar{t}$ and W + jets & $4-8$ & $15$ \\ \hline
$t \bar{t}$ semi-leptonic & $5$ & $22$ \\
$t \bar{t}$ di-leptonic & $10$ & $20$ \\ \hline
\hline
\end{tabular}
\caption{Expected statistical and systematic uncertainties on
the background, as expected by ATLAS in 1 fb$^{-1}$. The top rows
apply to the no-lepton mode, the bottom rows to the one-lepton
mode, and $t \bar{t}$ and W + jets apply to both modes.}
\label{tab:bg}
\end{table}

\section{Models and benchmarks}

Obviously, the main objective for CMS and ATLAS in searches for new physics
is: don't miss it~\cite{autermann}. Therefore, signatures and topologies
must be covered as complete and as general as possible.
The searches should be kept robust and inclusive, in
combination with exclusive measurements making use of specific
supersymmetry-related signatures.

The baseline searches are done assuming R-parity conservation,
where production and decay of squarks and gluinos lead to
energetic jets, $E_T^{\mathrm{miss}}$ from the unobserved lightest
supersymmetric particle, and possibly leptons. R-parity
violating searches are also performed, but are not covered here.
The interpretation of results is typically performed in the
mSUGRA framework, but also in the NUHM, GMSB and AMSB frameworks.

In order to study selection cuts, use is made of a number of
SUSY benchmark points with a specific choice of model parameters.
Most important is the coverage of signatures and a good representation
of phase space; the exact details of each point are not very
relevant. CMS and ATLAS have each chosen their own set of
benchmark points, for ease of comparison Table~\ref{tab:bm} gives
a rough translation of mSUGRA benchmark points in use.

\begin{table}
\begin{tabular}{|llll|l|}
\hline
Snowmass & CMS & ATLAS & BDEGMOPW &  Description \\
\hline
SPS 1a' & LM1 & SU3 & B' & ``bulk'' \\
SPS 1b  &     & SU8.1 & &  ``high $\tan \beta$ bulk'' \\
SPS 2   & LM9 &     & E' & ``focus point'' \\
        & LM7 & SU2 &    & ``high $m_0$ focus point'' \\
SPS 3   & LM6 & SU1 & C' & ``co-annihilation'' \\
SPS 4   & LM2 & SU6 & I'/L' & ``high $\tan \beta$'' \\
        &     & SU4 &    & ``low mass'' \\
\hline
\end{tabular}
\caption{A rough comparison of mSUGRA benchmark points in use
by CMS and ATLAS, and the Snowmass~\cite{snowmass} and 
BDEGMOPW~\cite{ellisetal} points. 
}
\label{tab:bm}
\end{table}

\section{Inclusive searches}

\subsubsection{Jets plus $E_T^{\mathrm{miss}}$ plus X}
\label{sec:msugrasearch}


In order to study the selection and discovery reach, recent ATLAS studies have
assumed a data sample of 1 fb$^{-1}$, and a corresponding knowledge of
detector-related systematic uncertainties. Furthermore, the expected uncertainties
of the background, as derived with data-driven estimation methods discussed
earlier, are taken into account. 
The significance of the signal is then calculated from the probability
of the background, including uncertainties, to fluctuate to the signal.

ATLAS makes a baseline selection of at least four energetic jets,
significant $E_T^{\mathrm{miss}}$, and transverse sphericity 
$S_T > 0.2$~\cite{autermann}. In the no-lepton search mode, no
isolated high $p_T$ electron or muon is allowed.
A discriminating variable between SUSY and background
is the effective mass $M_{\mathrm{eff}}$, defined as the sum of the $p_T$
of the leading four jets and $E_T^{\mathrm{miss}}$.
With an additional cut on $M_{\mathrm{eff}} > 800$ GeV,
a sensitivity of (significantly) more than $5 \sigma$ is reached for 
all ATLAS benchmark points except SU2. Analyses optimized for at least two or 
at least three jets have somewhat better significance, but are more
sensitive to the QCD multi-jet background. 

In the one-lepton mode, one identified high $p_T$ isolated electron
or muon is required, and 
the transverse mass of lepton and $E_T^{\mathrm{miss}}$ should be
larger than 100 GeV. The lepton requirement reduces the signal
somewhat, but suppresses QCD background, and again gives excellent
sensitivity.

The CMS studies have assumed either 1 or 10 fb$^{-1}$, and a
corresponding knowledge of detector-related systematic uncertainties.
In the no-lepton mode,
CMS demands at least three energetic central jets and large $E_T^{\mathrm{miss}}$,
and vetoes isolated leptons.
Finally, $H_T > 500$ GeV is demanded, where $H_T$ is the sum of the $E_T$ of
jets 2, 3 and 4, and $E_T^{\mathrm{miss}}$.
Similar cuts, but requiring an isolated lepton, are used on the one-lepton mode.
Further analyses look for tau leptons, a Z or a top quark together with
jets and $E_T^{\mathrm{miss}}$.

\begin{figure}[tb]
  \includegraphics[width=0.44\textwidth]{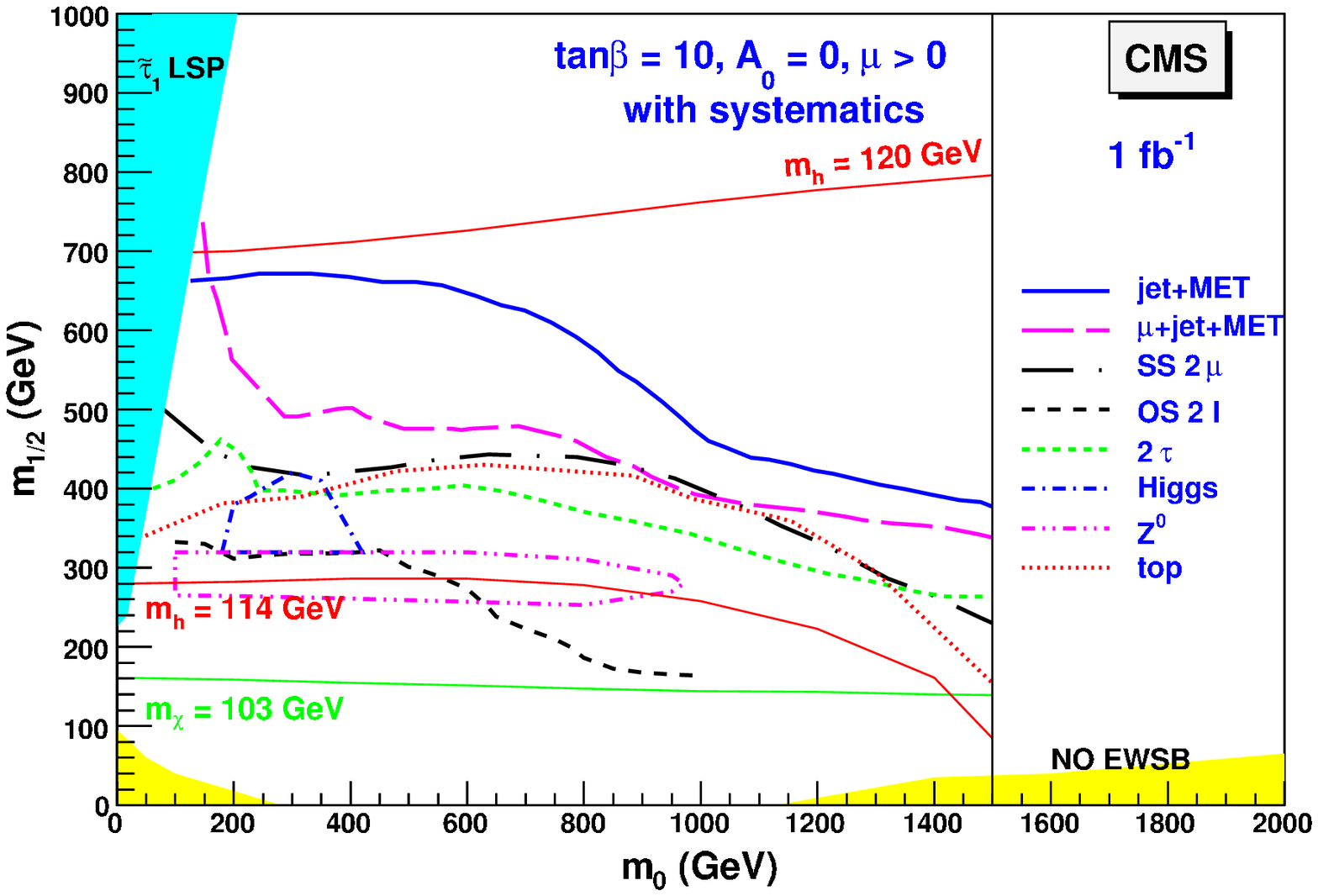}
  \includegraphics[width=0.37\textwidth]{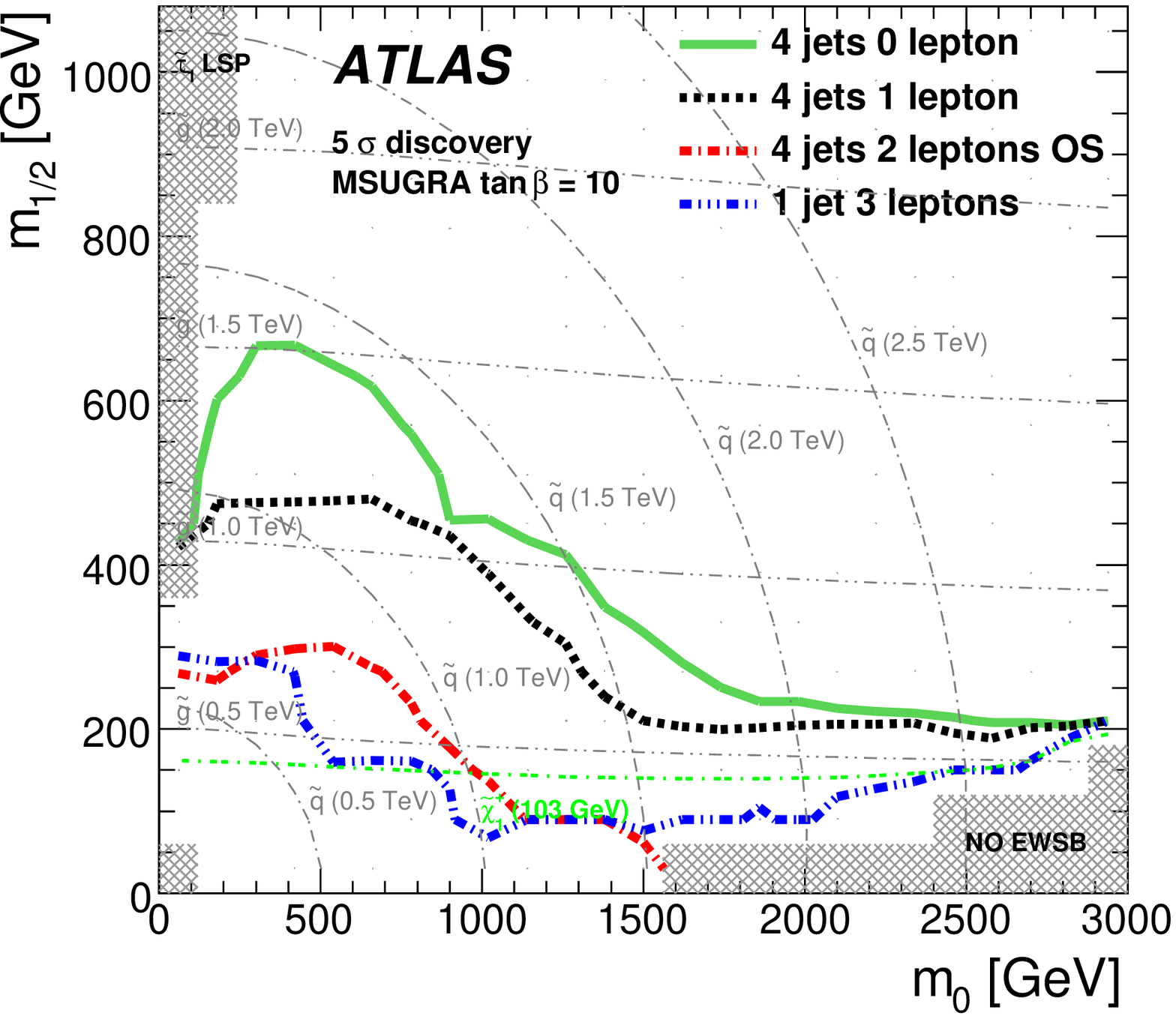} 
  \caption{Discovery reach of CMS (left) and ATLAS (right) in 
  $m_{1/2}$ and $m_0$ ($A_0 = 0$, $\tan \beta = 10$, $\mu > 0$) in
  mSUGRA, for 1 fb$^{-1}$.
  \label{fig:msugrareach}
}
\end{figure}

\begin{figure}[tb]
  \includegraphics[width=0.44\textwidth]{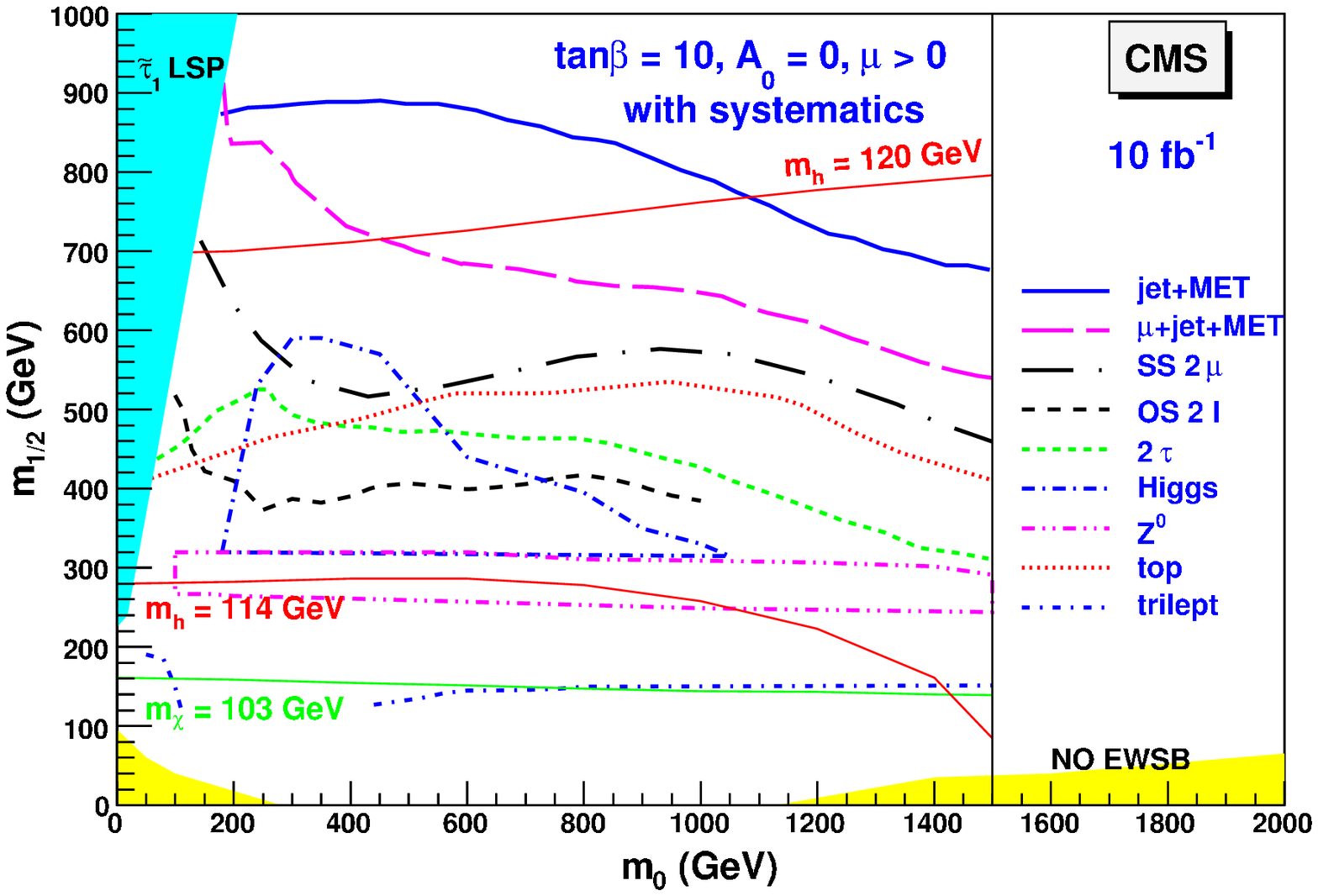}
  \includegraphics[width=0.37\textwidth]{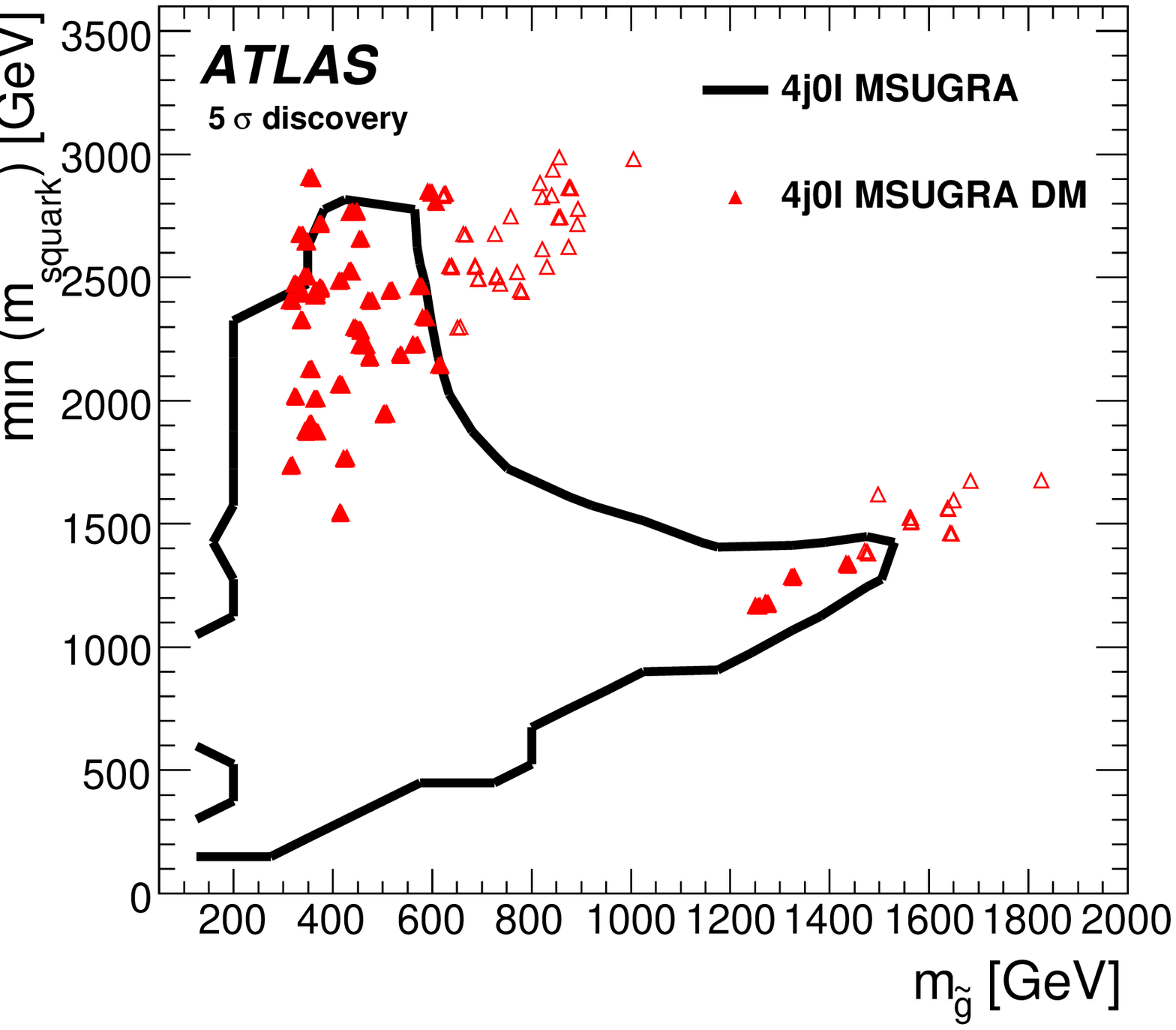}
  \caption{Idem for CMS in 10 fb$^{-1}$, and for ATLAS in
  1 fb$^{-1}$ expressed in squark and gluino masses.
  \label{fig:moremsugrareach}
}
\end{figure}

The CMS and ATLAS discovery reach for 1 fb$^{-1}$, interpreted in the mSUGRA model
for $A_0 = 0$, $\tan \beta = 10$, $\mu > 0$, is shown in 
Figure~\ref{fig:msugrareach}; clearly these sensitivities are comparable.
Figure~\ref{fig:moremsugrareach} shows the expected CMS reach in 10 fb$^{-1}$
and the ATLAS reach for 1 fb$^{-1}$ expressed in squark and gluino mass range.


Further interesting search modes include the di-lepton 
and tri-lepton searches. Demanding the two leptons in the
di-lepton mode to have the same charge effectively suppresses
background, but leaves a significant SUSY signal~\cite{pakhotin}. 
The tri-lepton
searches are sensitive to direct chargino/neutralino production,
for example in the ``focus region'', but need significant
luminosity ($> 10$ fb$^{-1}$)~\cite{tapas}. In these modes, the
background can probably be estimated from data with similar
methods as those used in the one-lepton mode.

\subsubsection{GMSB models}

In models with gauge-mediated SUSY breaking (GMSB), the LSP is
a gravitino that can be very light. The next-to-lightest SUSY
particle (NLSP) is typically either the lightest neutralino or
a stau (or the various sleptons are almost degenerate co-NLSPs).
The NLSPs decay into the LSP plus typically a photon, or a
lepton (tau). This decay can be prompt, or the NLSP can have
a significant lifetime, decaying away from the primary vertex or
even leaving the detector.

\begin{figure}[tb]
  \includegraphics[width=0.47\textwidth]{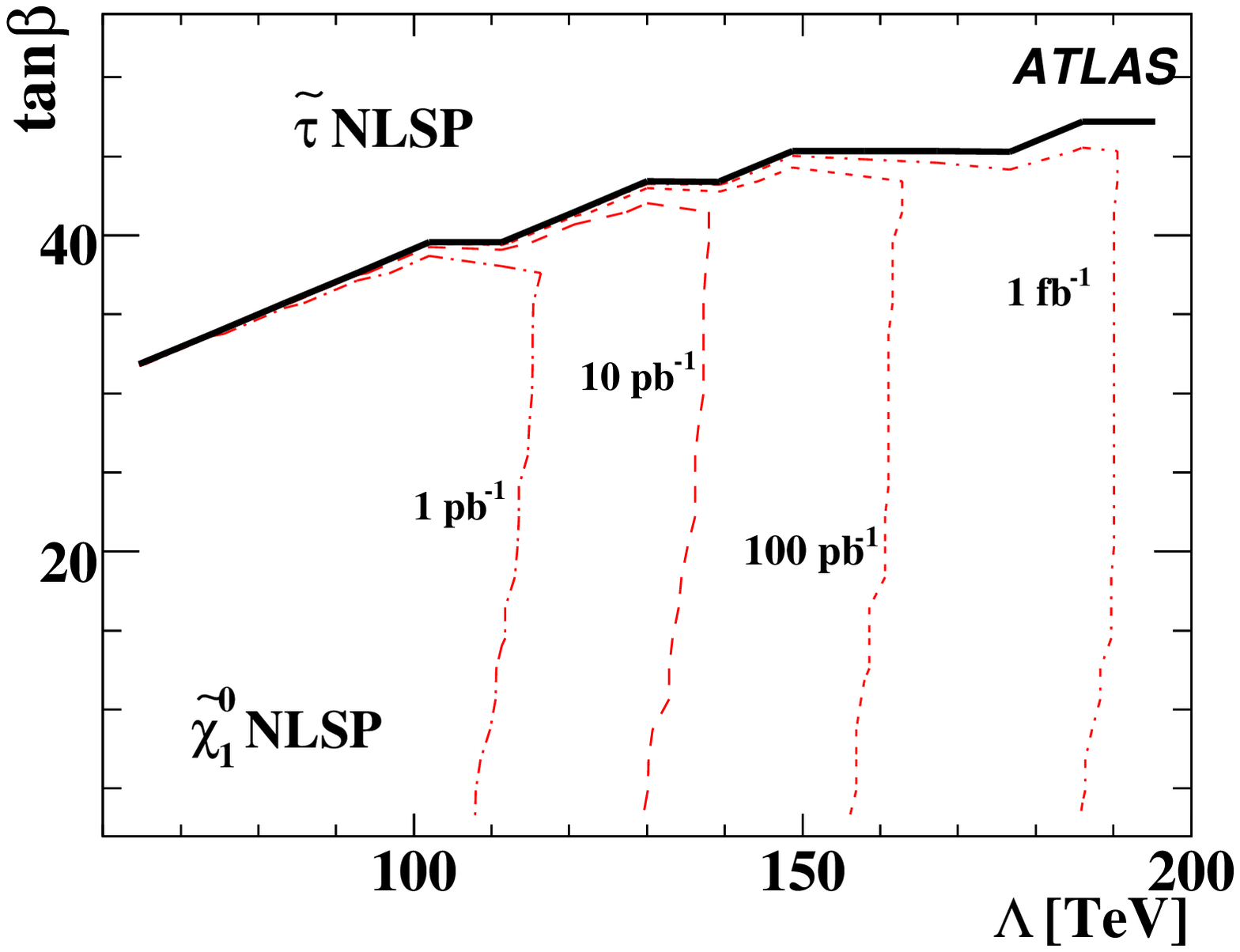}
  \includegraphics[width=0.34\textwidth]{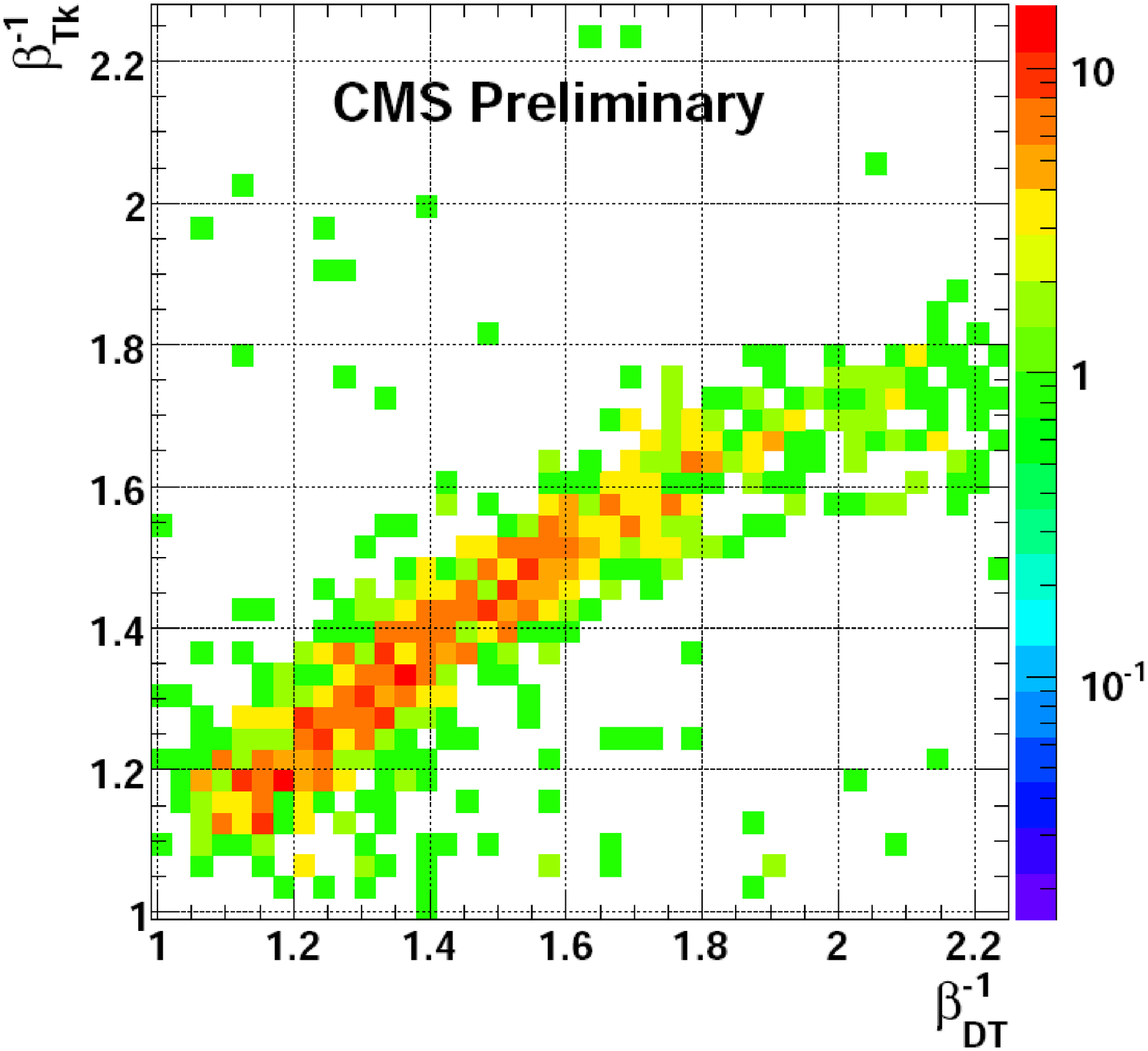} 
  \caption{Left: ATLAS discovery reach in GMSB parameters
$\Lambda$ and $\tan \beta$ in the analysis that searches for isolated
high $p_T$ photons and $E_T^{\mathrm{miss}}$. 
Right: $1/\beta$ measured from the CMS tracker (Tk) versus $1/\beta$
measured from the CMS muon drift tubes (DT), for a long-living stop quark with a 
mass of 500 GeV. The SM backgrounds cluster around
$1/\beta_{\mathrm{Tk}} = 1$ and/or $1/\beta_{\mathrm{DT}} = 1$.
  \label{fig:photonstudies}
}
\end{figure}

The ATLAS strategy for a search for prompt energetic photons
from $\tilde{\chi}^0_1 \to \gamma \tilde{G}$ leads to a sensitivity
as shown in Figure~\ref{fig:photonstudies} (left).

Neutralinos decaying away from the main vertex can lead to
photons in the detector that do not point to the main vertex.
Special reconstruction techniques must be applied to reconstruct
such photons, and both CMS and ATLAS have been studying this.

\begin{figure}[tb]
  \includegraphics[width=0.28\textwidth]{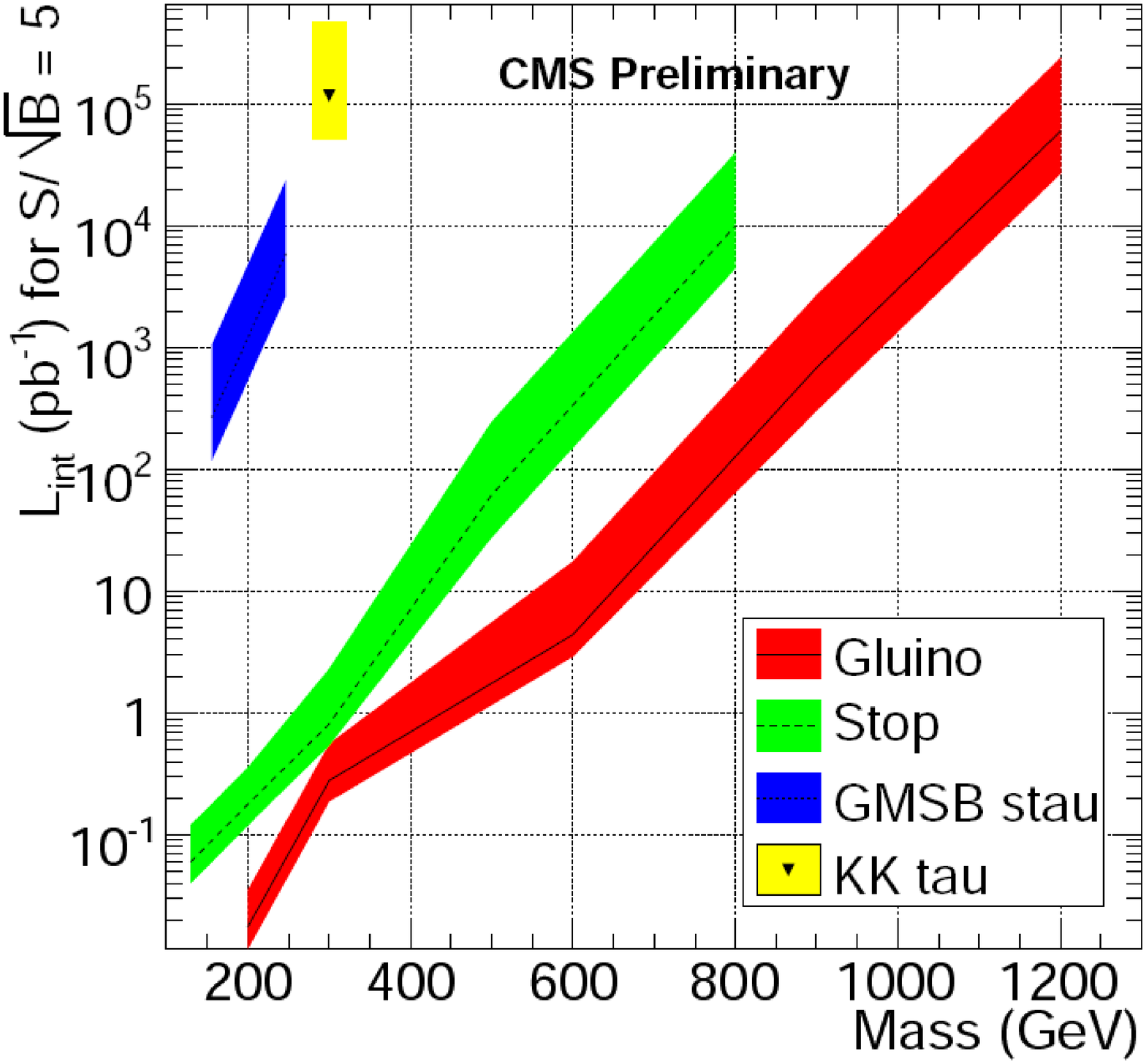}
  \includegraphics[width=0.36\textwidth]{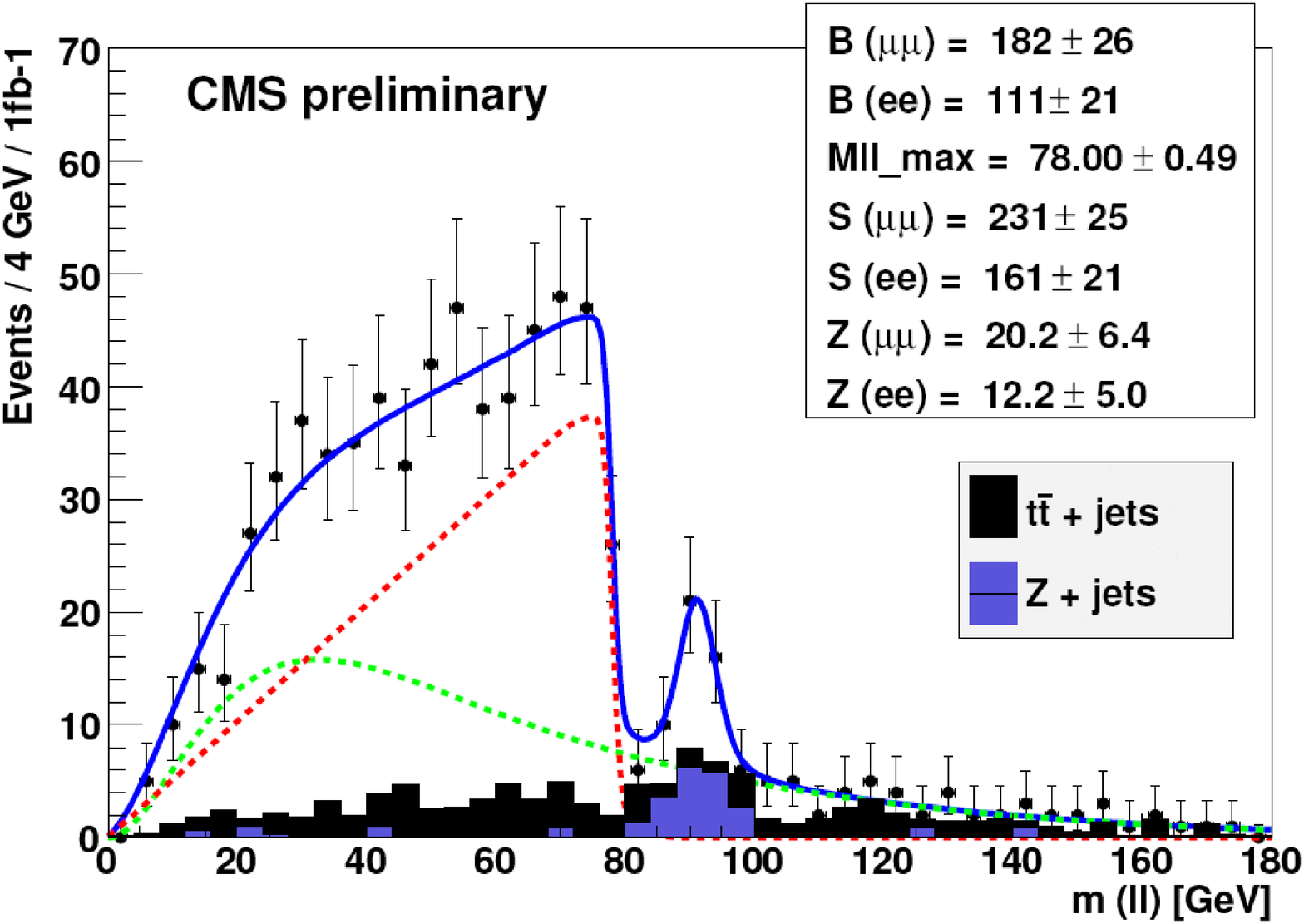}
  \caption{Left: luminosity needed by CMS to make a 5$\sigma$ discovery
of several classes of massive semi-stable charged particles as a function 
of their mass. Right: The edge in dilepton mass in the CMS LM1 model,
for 1 fb$^{-1}$. Red: signal pdf, green: flavour-symmetric background pdf.
  \label{fig:otherstudies}
}
\end{figure}

Massive charged particles with a long lifetime, for
example long-living stau leptons or stop quarks, will
leave a track in the tracking detectors of CMS or ATLAS. A decay
inside the tracking detectors can give rise to tracks with
significant kinks. NLSPs with an even longer lifetime leave a 
track through the whole detector, from tracking detectors to
the muon chambers. Such a track is thus reconstructed as a muon;
however due to the large NLSP mass its velocity $\beta$ might be
significantly smaller than $c$~\cite{teyssier}. Such slow particles may pose a
problem to trigger and reconstruction. Both CMS and ATLAS
have studied this~\cite{cmsexo,atlascsc}. Figure~\ref{fig:photonstudies} (right)
shows for CMS the measured $1/\beta$ from the tracker versus the
measured $1/\beta$ from the muon system for a long-living stop quark 
with a mass of 500 GeV. If both momentum and velocity can be reconstructed,
the NLSP mass can be estimated. 
Figure~\ref{fig:otherstudies} (left) shows the amount of luminosity needed
by CMS to make a 5$\sigma$ discovery of several classes of massive semi-stable 
charged particles as a function of their mass.

\section{Interpretation}

An inclusive discovery of new physics with jets and $E_T^{\mathrm{miss}}$,
or with signatures as in the GMSB model, would be spectacular,
but would not reveal much of the underlying physics.
Several models of new physics can fit the observed data, and certainly
the new physics can also be of a kind outside any existing model.
Establishing the nature of any observed deviation from the SM demands
hard work and ingenuity. Even then, ambiguities may remain.

Interpretation of any excess in terms of SUSY asks for consistency
of signals in various final states, mass scales, branching fractions,
cross-sections, and a proof of the spin of the newly observed particles.
 A first estimate
of the mass scale of squarks and gluinos can be derived from the
effective mass distribution and from the measured cross-sections.
More information, however, can be gained from exclusive studies.

In particular, one can try and select a suitable decay chain like
\begin{equation}
\tilde{q}_L \to \tilde{\chi}^0_2 q (\to \tilde{\ell}^{\pm} \ell^{\mp} q)
\to \tilde{\chi}^0_1 \ell^+ \ell^- q ,
\end{equation}
and measure invariant masses of combinations of objects like
the two leptons, the leptons and the jet, or one lepton and the jet.
The distributions of these masses typically have thresholds and
edges sensitive to the masses of particles in the decay chain.
Figure~\ref{fig:otherstudies} (right) shows a study of the
dilepton mass edge by CMS in the LM1 benchmark model, in 1 
fb$^{-1}$~\cite{cmsnewnote}.

ATLAS has studied the precision of parameter extraction in the
SU3 and SU4 models with 1, or 0.5 fb$^{-1}$~\cite{wienemann}. Already such limited
luminosity will give a first hint of underlying parameters.

A clear hint that SUSY particles are being produced could come
from a measurement of particle spin. This is difficult, and has
been studied by ATLAS for neutralinos and sleptons~\cite{spin}.



\section{Flavour-oriented studies}

Since in SUSY flavour-universality is very likely broken, it is
interesting to perform studies that concentrate on the properties
of individual flavours.


A study of the mass distribution of $e^{\pm} \mu^{\mp}$ combinations
is sensitive to lepton-flavour violating neutralino decays
$\tilde{\chi}_2^0 \to e^{\pm} \mu^{\mp} \tilde{\chi}_1^0$. A CMS
study has shown $5 \sigma$ sensitivity to lepton-flavour violating branching
fractions at the 4\% level with 10 fb$^{-1}$.

The lightest stop quark, $\tilde{t}_1$, is likely to be the lightest
squark. 
CMS has analyzed the potential of a stop quark discovery through its
decay $\tilde{t} \to t \tilde{\chi}_2^0 \to t \tilde{\ell}_R \ell
\to t \ell \ell \tilde{\chi}_1^0$.
In the events, the top quarks are kinematically reconstructed with
a kinematic fit. In the LM1 point, $M_{\tilde{t}_1} \approx 400$ GeV,
and a $5 \sigma$ discovery is possible with 200 pb$^{-1}$.

ATLAS has studied $\tilde{g} \to \tilde{t} t \to \tilde{\chi}^{\pm} b t$
in the SU4 (low mass) point. In this point, $M_{\tilde{t}_1} \approx 
200$ GeV, and it is possible to reconstruct the events and plot the
top-bottom mass distribution, which shows an edge sensitive to the
stop mass. In a sample of 200 pb$^{-1}$, almost 1000 signal events
would be observed, against 100 SM background events.

Furthermore, ATLAS has studied a scenario in which the lightest stop
is lighter than the top quark. Top-quark pair production is the largest
background to this search, but can be subtracted using a side-band
technique. Approximately 1 fb$^{-1}$ of data should be enough to see
a signal.


\section{Conclusions and outlook}

We are at the threshold of exciting times. 
Some 15 years of planning,
designing, construction, and installation of the detectors
are coming to an end. 
With the LHC data we will either
discover SUSY, or push its mass scale so high that it is no longer
a natural solution to the fine tuning problem. Let the data decide.

Excellent prospects for the SUSY search already exist with little
luminosity. However, what CMS and ATLAS really need now is luminosity
to shake down the detector, the trigger, the reconstruction, and
understanding of the backgrounds. Only data can give the confidence
that is needed to claim evidence for new phenomena. 
With 1-2 fb$^{-1}$
of integrated luminosity (2009?),
we will be able to gain that confidence, and we will be sensitive
to a very interesting region of SUSY parameter space~\cite{heinemeyer}.

\begin{theacknowledgments}
  The author wishes to thank everyone in CMS and ATLAS whose work
  contributed to this talk.
\end{theacknowledgments}






\end{document}